\documentclass[10pt]{iopart}
\usepackage{graphicx}
\usepackage{cite}
\begin{document}

\paper{Smearing of the phase transition in Ising systems with planar defects}

\author{Thomas Vojta}

\address{Department of Physics, University of Missouri - Rolla, Rolla, MO 65409, USA}

\begin{abstract}
We show that phase transitions in Ising systems with planar defects, i.e., disorder perfectly
correlated in two dimensions are destroyed by smearing. This is caused by effects similar to but
stronger than the Griffiths phenomena: Exponentially rare spatial regions can develop true static
long-range order even when the bulk system is still in its disordered phase. Close to the smeared
transition, the order parameter is very inhomogeneous in space, with the thermodynamic (average)
order parameter depending exponentially on temperature. We determine the behavior using extremal
statistics, and we illustrate the results by computer simulations.
\end{abstract}



\section{Introduction}
\label{sec:intro}

Phase transitions in random systems remain one of the important and only partially solved problems
in statistical physics. Originally, it was thought that even weak disorder destroys any critical
point because the disordered system divides itself up into spatial regions which undergo the
transition at different temperatures, leading to a smeared transition in which there would be no
sharp singularities in thermodynamic quantities (see \cite{Grinstein} and references therein).
However, it was soon found that classical continuous phase transitions are generically sharp in
the presence of weak, short-range correlated disorder. If a clean critical fixed point (FP)
fulfills the Harris criterion \cite{Harris74} $\nu\ge 2/d$, where $\nu$ is the correlation length
exponent and $d$ is the spatial dimensionality, the disorder decreases under coarse graining. The
system becomes asymptotically homogeneous at large length scales. Thus, the clean critical point
is perturbatively stable against disorder, and the critical behavior of the random system is
identical to that of the corresponding clean system. Macroscopic observables are self-averaging at
the critical point, i.e., the relative width of their probability distributions goes to zero in
the thermodynamic limit \cite{AharonyHarris96,WisemanDomany98}.

Even if the Harris criterion is violated, the transition will generically remain sharp, but the
critical behavior will be different from that of the clean system. Depending on the fate of the
disorder under coarse graining it can either be of finite or of infinite disorder type. In the
first case the system remains inhomogeneous at all length scales with the relative strength of the
inhomogeneities approaching a finite value for large length scales. These transitions are
controlled by renormalization group (RG) fixed points with finite disorder. Macroscopic
observables are not self-averaging, the relative width of their probability distributions
approaches a size-independent constant \cite{AharonyHarris96,WisemanDomany98}. Examples of
critical points in this class include the dilute three-dimensional Ising model or classical spin
glasses. The second case occurs when the relative magnitude of the inhomogeneities increases
without limit under coarse graining. The corresponding fixed points are called infinite-disorder
fixed points. Here, the probability distributions of macroscopic variables become very broad (on a
logarithmic scale) with the width increasing with system size. This can lead to activated rather
than the usual power-law scaling at the critical point. A famous classical example of an
infinite-disorder critical point occurs in the McCoy-Wu model \cite{McCoyWu}.

In recent years, another aspect of phase transitions in disordered systems, the Griffiths
phenomena, has regained a lot of attention. Griffiths phenomena are non-perturbative effects of
rare disorder fluctuations in the vicinity of a phase transition. They can be understood as
follows:  Disorder in general decreases the critical temperature $T_c$ from its clean value
$T_c^0$. In the temperature region $T_c<T<T_c^0$ the system does not display global order.
However, in an infinite system one will find arbitrarily large regions that are devoid of
impurities with a small but non-zero probability. These rare regions or 'Griffiths islands' are
locally in the ordered phase while the bulk system is still in the disordered phase. Their
fluctuations are very slow because flipping them requires changing the order parameter in a large
volume. Griffiths \cite{Griffiths} was the first to show that this leads to a non-analytic free
energy everywhere in the region $T_c<T<T_c^0$, which is now known as the Griffiths phase
\cite{Randeria} or the Griffiths region. In generic classical systems, the contribution of the
Griffiths singularities to thermodynamic (equilibrium) observables is very weak since the
singularity in the free energy is only an essential one \cite{Griffiths,Bray89}. In contrast, the
consequences for the dynamics are much more severe with the rare regions dominating the behavior
for long times. In the  Griffiths region, the spin autocorrelation function $C(t)$ decays very
slowly with time $t$, like $\ln C(t) \sim -(\ln t)^{d/(d-1)}$ for Ising systems
\cite{Randeria,Dhar,Dhar88,Bray88a,Bray88b}, and like $\ln C(t) \sim -t^{1/2}$ for Heisenberg
systems \cite{Bray88a,Bray87}. More recently, these results have been confirmed by rigorous
calculations for the equilibrium \cite{Dreyfus,Gielis95} and dynamic \cite{Cesi97} properties of
disordered Ising models.

In many  real systems, the disorder is not generated by point-like defects but by dislocations or
disordered layers or grain boundaries. Often, extended impurities in a $d$-dimensional system can
be modeled by disorder perfectly correlated in $d_C=$ 1 or 2 dimensions but uncorrelated in the
remaining $d_\bot=d-d_C$ dimensions. While it is generally accepted that long-range disorder
correlations increase the effects of the impurities, their influence on the phase transition has
been controversial. Early RG work \cite{Lubensky} based on a single expansion in $\epsilon=4-d$
did not produce a critical fixed point. This was interpreted as a smeared transition or a first
order one \cite{Rudnick,Andelman}. However, an additional expansion in the number $d_C$ of
correlated dimensions \cite{Dorogovtsev,BoyCardy,DeCesare} cured this problem and gave rise to
critical fixed points with conventional power-law scaling.  Further support for the sharp
transition scenario came from Monte-Carlo simulations of a 3D Ising model with planar defects
\cite{LeeGibbs}. All the perturbative RG studies miss, however, the effects of rare regions and
Griffiths phenomena. These effects have been studied in detail in the already mentioned McCoy-Wu
model \cite{McCoyWu}, a disordered 2D Ising model in which the disorder is perfectly correlated in
one dimension and uncorrelated in the other. The transition in this model was originally thought
to be smeared. It was later found that it is sharp but of infinite-disorder type
\cite{McCoy69,dsf9295}. Similar behavior has been found in several random quantum systems which
are related to classical systems with linear defects
\cite{YoungRieger96,Pich98,Motrunich00,ma79,bhattlee82,fisher94}. Based on these results it was
believed for a long time that phase transitions generically remain sharp even in the presence of
extended disorder.

In this paper we show, however, that this belief is not generally true. Specifically, we show that
for Ising order parameter symmetry, planar defects destroy the sharp continuous phase transition
via effects which are similar to but stronger than the usual Griffiths phenomena: In systems with
planar defects true {\em static long-range order} can develop on isolated rare regions. As a
result, the order parameter is very inhomogeneous in space close to the smeared transition, and
the thermodynamic order parameter depends exponentially on temperature. We also show that this
disorder-induced smearing generically occurs in an entire class of transitions in systems with
extended disorder. The paper is organized as follows. In Sec.\ \ref{sec:rounding}, the model is
introduced, and the mechanism for the smearing of the transition is explained. In Sec.\
\ref{sec:lifshitz} we use Lifshitz tail arguments to determine the behavior in the 'tail' of the
rounded transition, i.e., for a very low concentration of ordered islands. Section
\ref{sec:numerics} is devoted to a numerical study of a three-dimensional infinite-range Ising
model which illustrates the smearing of the phase transition. Finally, in Sec.\
\ref{sec:discussion} we discuss the generality of the disorder-induced smearing, the relation to
Griffiths phenomena, favorable conditions for observing the smearing, and the influence of the
order parameter symmetry.

\section{Rare regions, inhomogeneous order, and smearing}
\label{sec:rounding}

For definiteness we consider a $d$-dimensional $\phi^4$ theory with random-$T_c$ type disorder
completely correlated in the $d_C=2$ directions $x_1$ and $x_2$ but uncorrelated in the remaining
$d_\bot=d-d_C$ directions $x_3,\dots,x_d$. The action is given by
\begin{equation}
S = \int d^d r ~ \phi({\bf r})\left[ t+\delta t ({\bf r_\bot}) -\partial^2_{\bf r} \right]
\phi({\bf r}) + u\int d^d r ~\phi^4({\bf r})~. \label{eq:action}
\end{equation}
Here $\phi({\bf r})$ is a scalar field, $t$ is the dimensionless distance from the clean critical
point and $\delta t({\bf r_\bot})$ introduces the quenched disorder. ${\bf r_\bot}$ is the
projection of ${\bf r}$ on the uncorrelated directions $x_3,\dots,x_d$. We consider two different
types of disorder. In the first type, Poisson (or dilution) disorder, $\delta t({\bf r_\bot})$ is
given by
\begin{equation}
\delta t({\bf r_\bot}) = \sum_i V[{\bf r_\bot}-{\bf r_\bot}(i)]
\end{equation}
where ${\bf r_\bot}(i)$ are the random positions of planar impurities of spatial density $c$, and
$V({\bf r_\bot})$ is a non-negative short-ranged impurity potential. The second type of disorder
is a Gaussian distribution with zero mean and a correlation function
\begin{equation}
\langle \delta t({\bf r_\bot})\delta t({\bf r_\bot}')\rangle = \Delta^2 \delta({\bf r_\bot}-{\bf
r_\bot}')~.
\end{equation}

Now consider the effects of rare disorder fluctuations in this system: Analogous to the Griffiths
phenomena, there is a small but finite probability for finding large spatial regions in ${\bf
r}_\bot$ direction which are more strongly coupled than the bulk system. For Poisson disorder,
these are regions devoid of impurities while for Gaussian disorder it would be regions with a
negative average $\langle \delta t({\bf r_\bot}) \rangle$. These rare regions can be locally in
the ordered phase, even if the bulk system is still in the disordered phase. For Poisson disorder
which fulfills $\delta t({\bf r_\bot})>0$ this starts to happen right below the clean transition,
i.e., for $t<0$ \cite{huse_private} while for the unbounded Gaussian disorder, locally ordered
rare regions exist for all $t$. Since the defects in our system are planar, the rare regions are
infinite in the $x_1$ and $x_2$ directions but finite in the remaining directions. This is a
crucial difference to systems with uncorrelated disorder, where the rare regions are finite
objects (see Sec.\ \ref{sec:discussion}).  In our system, each rare region is therefore equivalent
to a two-dimensional Ising system and can undergo a real phase transition {\em independently} of
the rest of the system. Thus, for planar defects, those rare regions which are in the ordered
phase will develop a non-zero static order parameter which can be aligned by an infinitesimally
small interaction or an infinitesimally small field.

The resulting phase transition is very different from conventional continuous phase transitions,
be it transitions controlled by clean or finite-disorder  or even infinite-randomness fixed
points. In all of those cases, which represent ``sharp'' phase transitions, a non-zero order
parameter develops as a collective effect of the entire system, accompanied by a diverging
correlation length in all directions. In contrast, in a system with planar defects, the order
parameter develops very inhomogeneously in space with different parts of the system (i.e.,
different ${\bf r}_\bot$ regions) ordering independently at different $t$. Correspondingly, the
correlation length in ${\bf r}_\bot$ direction remains finite for all temperatures. This defines a
smeared transition. We thus conclude that planar defects lead to a smearing of the phase
transition.

Above, we have defined a smeared continuous phase transition via the behavior of correlation
length. We now characterize it via the behavior of the free energy density or other global
thermodynamic variables.\footnote{Note that global thermodynamic variables (which average over the
entire sample) do not resolve the strong spatial inhomogeneities which are the hallmark of a
smeared transition, and thus do not provide complete information about the transition} Two
qualitatively different cases of smeared transitions must be distinguished. If the disorder
distribution is unbounded in the sense that it permits regions with a local $T_c=\infty$ (system
(\ref{eq:action}) with Gaussian disorder falls into this class), the total (average) order
parameter is non-zero for all temperatures, and the free energy density is analytic. In contrast,
if the disorder distribution is bounded, there is a true paramagnetic phase with zero order
parameter at high temperatures. At some temperature $T_0$, a non-zero order parameter starts to
develop on rare spatial regions, accompanied by an essential singularity in the free energy
density (which stems from the probability for finding a rare region). This singularity is weaker
than the power-law singularities at the sharp transitions mentioned above. The model
(\ref{eq:action}) with Poisson disorder falls into this second class, with $T_0$ being identical
to the critical temperature of the pure system.

\section{Lifshitz tail arguments}
\label{sec:lifshitz}

In this section we use extremal statistics to derive the leading thermodynamic behavior in the
'tail' of the smeared transition, i.e., in the parameter region where a few islands have developed
static order but their density is so small that they can be treated as independent. The approach
is similar to that of Lifshitz \cite{Lifshitz} and others for the description of the tails in the
electronic density of states. We first consider the Poisson type of disorder and later describe
the differences for Gaussian disorder.

\subsection{Poisson disorder}

The probability $w$ of finding a large region of linear size $L_R$ (in ${\bf r}_\bot$-space)
devoid of any impurities is, up to pre-exponential factors, given by
\begin{equation}
w \sim \exp( -c L_R^{d_\bot})~. \label{eq:wLR}
\end{equation}
As discussed in Sec.\ \ref{sec:rounding}, such a rare region develops static long-range order at
some $t_c(L_R)$ below the clean critical point $t_c^0=0$.\footnote{We use the bare value, $t_c^0
=0$, for the clean critical reduced temperature. For our purpose, the distinction between the bare
and the renormalized $t$ is of no importance.} The value of $t_c(L_R)$ varies with the size of the
region: The largest islands develop long-range order closest to the clean critical point. Finite
size scaling (for the clean system, because the island is devoid of impurities) yields
\begin{equation}
t_c^0 - t_c(L_R) =|t_c(L_R)| = A \,L_R^{-\phi} \label{eq:FSS}
\end{equation}
where $\phi$ is the finite-size scaling shift exponent and $A$ is the amplitude for the crossover
from $d$ dimensions to a slab geometry infinite in two dimensions but finite in $d_\bot=d-2$
dimensions. Combining (\ref{eq:wLR}) and (\ref{eq:FSS}) we obtain the probability for finding a
rare region which becomes critical at $t_c$ as
\begin{equation}
w(t_c) \sim \exp  (-B ~|t_c|^{-d_\bot/\phi}) \qquad (\textrm{for } t\to 0-)
 \label{eq:w-dilute}
\end{equation}
where the constant $B$ is given by $B=c\,A^{d_\bot/\phi}$.  The total (or average) order parameter
$m$ is obtained by integrating over all rare regions which are ordered at $t$, i.e., all rare
regions having $t_c>t$. Since the functional dependence on $t$ of the order parameter on a given
island is of power-law type it does not enter the leading exponentials but only the
pre-exponential factors. Therefore we obtain to exponential accuracy
\begin{eqnarray}
m \sim \exp  (-B ~|t_c|^{-d_\bot/\phi}) \qquad (\textrm{for } t\to 0-)~. \label{eq:m-dilute}
\end{eqnarray}
The homogeneous magnetic susceptibility consists of two different contributions, one from the
islands on the verge of ordering and one from the bulk system still deep in the disordered phase.
The bulk system provides a finite, noncritical background susceptibility throughout the tail
region of the smeared transition. In order to estimate the contribution of the islands consider
the onset of local ordering at the clean critical temperature $t_c^0=0$. Using (\ref{eq:w-dilute})
for the density of the islands we can estimate
\begin{equation}
\chi \sim \int_0^\Lambda dt  ~t^{-\gamma}~\exp  (-B ~t^{-d_\bot/\phi})
\label{eq:chi}
\end{equation}
which is finite because the exponentially decreasing island density overcomes the power-law
divergence of the susceptibility of an individual island. Here $\gamma$ is the clean
susceptibility exponent and $\Lambda$ is related to a lower cutoff for the island size. Once
ordered islands exist they produce an effective background magnetic field everywhere in space
which cuts off any possible divergence. Therefore we conclude that the homogeneous magnetic
susceptibility does not diverge anywhere in the tail region of the smeared transition. However,
there is an essential singularity at the clean critical point produced by the vanishing density of
ordered islands.

The spatial magnetization distribution in the tail of the smeared transition is very
inhomogeneous. On the already ordered islands, the local order parameter $m({\bf r})$ is of the
same order of magnitude as in the clean system. Away from these islands it decays exponentially
with the distance from the nearest island. The probability distribution $P[\log m({\bf r})]$ will
therefore be very broad, ranging from $\log m({\bf r}) = O(1)$ on the largest islands to $\log
m({\bf r}) \to -\infty$ on sites very far away from an ordered island. The {\em typical} local
order parameter $m_{typ}$ can be estimated from the typical distance of any point to the nearest
ordered island. From (\ref{eq:w-dilute}) we obtain
\begin{equation}
r_{typ} \sim \exp  (B ~|t|^{-d_\bot/\phi}/d_\bot) ~. \label{eq:rtyp}
\end{equation}
At this distance from an ordered island, the local order parameter has decayed to
\begin{equation}
m_{typ} \sim e^{-r_{typ}/\xi_0} \sim \exp \left[ -C \exp(B ~|t|^{-d_\bot/\phi}/d_\bot)\right]~
\label{eq:mtyp}
\end{equation}
where $\xi_0$ is the bulk correlation length (which is finite and changes slowly throughout the
tail region of the smeared transition)  and $C$ is constant. A comparison with (\ref{eq:m-dilute})
gives the relation between $m_{typ}$ and the thermodynamic order parameter $m$,
\begin{equation}
|\log m_{typ}| \sim m^{-1/d_\bot} \label{eq:mtypav}~.
\end{equation}
Thus, $m_{typ}$ decays exponentially with $m$ indicating an extremely broad local order parameter
distribution. In order to determine the functional form of the local order parameter distribution,
first consider a situation with just a single ordered island at the origin of the coordinate
system. For large distances ${\bf r}$ the magnetization falls off exponentially like $m({\bf r}) =
m_0~ e^{-r_\bot/\xi_0}$. The probability distribution of $x=\log[m({\bf r})]=\log m_0
-r_\bot/\xi_0$ can be calculated from
\begin{equation}
P(|x|) = \left |\frac {dN}{dx} \right| = \frac{dN}{dr_\bot} \left | \frac {dr_\bot}{dx}\right |
=\xi_0 \frac{dN}{dr_\bot}
       \sim \xi_0 r_\bot^{d-1}
\end{equation}
where $dN$ is the number of sites at a distance from the origin between $r_\bot$ and
$r_\bot+dr_\bot$ or, equivalently, having a logarithm of the local magnetization between $x$ and
$x+dx$. For large distances we have $|x| \sim r_\bot$. Therefore, the probability distribution of
$\log m$ generated by a single ordered island takes the form
\begin{equation}
 P[\log(m)] \sim |\log(m)|^{d_\bot-1}   \qquad (\textrm{for } m \ll 1)~.
 \label{eq:plogm}
\end{equation}
In the tail region of the smeared transition the system consists of a few ordered islands whose
distance is large compared to $\xi_0$. The probability distribution of $\log[m({\bf r})]$ thus
takes the form (\ref{eq:plogm}) with a lower cutoff corresponding to the typical island-island
distance and an upper cutoff corresponding to a distance $\xi_0$ from an ordered island.

\subsection{Finite-size effects}
\label{subsec:FS}

It is important to distinguish effects of a finite size $L_C$ in the correlated directions and a
finite size $L_\bot$ in the uncorrelated directions. If $L_\bot$ is finite but $L_C$ is infinite
static order on the rare regions can still develop. In this case, the sample contains only a
finite number of islands of a certain size. As long as the number of relevant islands is large,
finite size-effects are governed by the central limit theorem. However, for $t\to 0-$ very large
and rare islands are responsible for the order parameter. The number $N$ of islands which order at
$t_c$ behaves like $N \sim L_\bot^{d_\bot} ~w(t_c)$. When $N$ becomes of order one, strong
sample-to-sample fluctuations arise. Using (\ref{eq:w-dilute}) for $w(t_c)$ we find that strong
sample to sample fluctuations start at
\begin{eqnarray}
|t_{L_\bot}| \sim \left( \frac {d_\bot} B \log (L_\bot) \right)^{-\phi/d_\bot}~.
\label{eq:tL-dilute}
\end{eqnarray}
Thus, finite size effects are suppressed only logarithmically.

Analogously, one can study the onset of static order in a sample of finite size $L_\bot$ (i.e.,
the ordering temperature of the largest rare region in this sample). For small sample size
$L_\bot$, the probability distribution $P(t_c)$ of the sample ordering temperatures $t_c$ will be
broad because some samples do not contain any large islands. With increasing sample size the
distribution becomes narrower and moves towards the clean $t_c^0$ because more samples contain
large islands. The maximum $t_c$ coincides with $t_c^0$ corresponding to a sample without
impurities. The lower cutoff corresponds to an island size so small that essentially every sample
contains at least one of them. Consequently, the width of the distribution of critical
temperatures in finite-size samples is governed by the same relation as the onset of the
fluctuations,
\begin{eqnarray}
\Delta t_{c} \sim \left ( \frac {d_\bot} B \log (L_\bot) \right)^{-\phi/d_\bot}~.  \label{eq:dtc}
\end{eqnarray}

We now turn to finite-size effects produced by a finite extension $L_C$ of the system in the
correlated directions.  In this case, the true static order on the rare regions is destroyed.
Whether or not the system develops long-range order depends on the number $d_\bot$ of uncorrelated
dimensions. If $d_\bot=1$ no static long-range order can develop, i.e. the transition is rounded
by conventional finite size effects {\em in addition} to the disorder-induced smearing. In
contrast, for $d_\bot>1$ a true phase transition is possible but requires a finite interaction
between the islands of the order of the temperature. This restores a {\em sharp} phase transition
at a reduced temperature $t_c (L_C)$. To estimate the relation between $L_C$ and $t_c (L_C)$ we
note that the interaction between two planar rare regions of linear size $L_C$ is proportional to
$L_C^2$ and decays exponentially with their spatial distance $r$, $E_{int} \sim L_C^2
\exp(-r/\xi_0)$, where $\xi_0$ is the bulk correlation length. With the typical distance given by
(\ref{eq:rtyp}) we obtain a double exponential dependence between $L_C$ and the critical reduced
temperature $t_c (L_C)$:
\begin{eqnarray}
\log( \log L_C) &\sim& |t_c|^{-d_\bot/\phi}~.
\end{eqnarray}

\subsection{Gaussian disorder}

In this subsection we apply Lifshitz tail arguments to the case of $\delta$-correlated Gaussian
disorder \cite{HalperinLax}. In contrast to the positive definite Poisson disorder, the Gaussian
disorder is unbounded for $\delta t({\bf r}) \to -\infty$. Therefore, locally ordered islands can
exist for all $t$, and, in principle, the tail region of the smeared transition stretches to $t
\to \infty$. In order to determine the probability for finding a rare region which orders at a
certain $t_c$, i.e., the Gaussian equivalent to (\ref{eq:w-dilute}), consider an island of linear
size $L_R$. The average disorder value $\langle \delta t \rangle$ in this region has the
distribution
\begin{equation}
P[\langle\delta t \rangle] \sim \exp \left[ -L_R^{d_\bot} \langle \delta t \rangle^2/(2\Delta)
\right]~.
\end{equation}
According to finite size scaling, an island of size $L_R$ with an average disorder value of
$\langle \delta t \rangle$ will develop static order at $t_c = -\langle \delta t \rangle - A\,
L_R^{-\phi}$. For fixed island size, the probability distribution of $t_c$ reads
\begin{equation}
P(t_c) \sim \exp \left[ -L_R^{d_\bot} (t_c + A\,L_R^{-\phi})^2/(2\Delta) \right]~. \label{eq:ptcL}
\end{equation}
We now calculate the stationary point of the exponent with respect to $L_R$ (saddle point
approximation) to determine which island size gives the dominating contribution. For
$(2-d_\bot/\phi)>0$, this leads to
\begin{equation}
(2-d_\bot/\phi)~A\,L_r^{-\phi} = t_c d/\phi~.
\end{equation}
This implies that small but very deep fluctuations are responsible for the tail of the smeared
transition (in contrast to Poisson disorder where the tail is produced by the largest islands).
Inserting into (\ref{eq:ptcL}) gives the desired probability as a function of $t_c$,
\begin{equation}
w(t_c) \sim  \exp ( -\bar B ~t_c^{2-d_\bot/\phi} ) \label{eq:w-Gauss}
\end{equation}
where $\bar B$ is a constant. In contrast, for $(2-d_\bot/\phi)<0$, the main contribution comes
from arbitrarily small islands. In this case the $\delta$-correlated disorder is unphysical and
should be replaced by a distribution with a finite correlation length. The resulting  $w(t_c)$ is
then purely Gaussian. In the following we will not consider this case.

Starting from (\ref{eq:w-Gauss}), one can derive the leading behavior in the tail of the smeared
transition ($t \to \infty$). Integrating over all ordered islands gives the total magnetization
\begin{equation}
m \sim  \exp ( -\bar B ~t_c^{2-d_\bot/\phi} )  ~. \label{eq:m-Gauss}
\end{equation}
The typical local magnetization $m_{typ}$ decays exponentially with the average magnetization $m$,
i.e.\ (\ref{eq:mtypav}) also holds in the Gaussian case as can be easily seen by starting the
derivation from (\ref{eq:w-Gauss}) rather than (\ref{eq:w-dilute}). For samples of finite size in
the uncorrelated directions the onset of sample-to-sample fluctuations and the width of the sample
ordering temperatures follow
\begin{equation}
 t_{L_\bot} \sim \Delta t_c \sim (\log L_\bot)^{1/(2-d_\bot/\phi)}~.
\end{equation}
If the samples are of finite size in the correlated directions but infinite in the uncorrelated
ones (and if $d_\bot>1$) a sharp phase transition is restored. The relation between $L_C$ and
$t_c$ is double-exponential:
\begin{equation}
 \log( \log L_C) \sim t_c^{2-d_\bot/\phi} ~.
\end{equation}
A comparison of the results for Gaussian and Poisson disorder shows that the functional form of
the thermodynamic relations at the smeared transition depends on the form of the disorder and is
thus not universal.

\section{Numerical results}
\label{sec:numerics}

\subsection{Model and method}

To illustrate the smearing of the phase transition we now show results from a numerical simulation
of a 3D Ising model with planar bond defects, i.e., $d_C=2$ and $d_\bot=1$. As discussed above,
the smearing of the phase transition is the result of exponentially rare events. Therefore large
system sizes are required to observe it. In order to reach sufficiently large sizes with a
reasonable effort and to retain the possibility of static order on an isolated rare region, we
consider a model with infinite-range interactions in the correlated directions (parallel to the
defects) but short-range interactions in the uncorrelated direction (perpendicular to the defect
planes). While the infinite-range model will not be quantitatively comparable to a short-range
model, it provides a simple example for the rounding mechanism introduced in Sec.\
\ref{sec:rounding}. Moreover, since the rounding mechanism only depends on the existence of static
order on the rare regions but not on any details, we expect that the results will qualitatively
valid for a short-range model, too (with the appropriate changes to the exponents in the relations
derived in Sec. \ref{sec:lifshitz}). Indeed, preliminary results \cite{Sknepnek} of Monte-Carlo
simulations of a 3D short-range Ising model with planar defects fully agree with this expectation.

The Hamiltonian considered in this section reads
\begin{equation}
\fl H= - \frac 1 {L_C^2}\sum_{x,y,z,y',z'}   S_{x,y,z} S_{x+1,y',z'}
   - \frac 1 {2 L_C^2}\sum_{{x,y,z,y',z'}} J_{x} S_{x,y,x} S_{x,y'z'}~ - h\sum_{x,y,z} S_{x,y,x} .
\label{eq:toy}
\end{equation}
Here $x,y,z$ are the integer-valued coordinates of the Ising spins. The magnetic field $h$ acts as
a symmetry breaker, it will be set to a small but non-zero value, typically $h=10^{-10}$. $L_C$ is
the system size in the $y$ and $z$ directions. The planar defects are parallel to the
$(y,z)$-plane. They are introduced via disorder in the in-plane coupling $J_{x}$ which is a
quenched binary random variable with the distribution $P(J) = (1-c)~ \delta(J-1) + c~
\delta(J-J_d)$ with $J_d<1$. Here, $c$ is the defect concentration and the deviation of $J_d$ from
1 measures the strength of the defects.  For our calculations we use $J_d=0$, i.e., very strong
defects. The fact that one can independently vary concentration and strength of the defects in an
easy way is the main advantage of this binary disorder distribution. However, it also has unwanted
consequences, viz. log-periodic oscillations of many observables as functions of the distance from
the critical point \cite{karevski}. These oscillations are special to the binary distribution and
unrelated to the smearing considered here; we will not discuss them further.

Because the interaction is infinite-ranged in the correlated ($y$ and $z$) directions, these
dimensions can be treated exactly: We introduce the plane magnetizations
\begin{equation}
m_x = \frac 1 {L_C^2} \sum_{y,z} S_{x,y,z}
\end{equation}
into the partition function and integrate out the original Ising variables $S_{x,y,z}$. In the
thermodynamic limit, $L_C\to\infty$, the resulting integral can be solved using saddle point
approximation. This leads to a set of coupled local mean-field equations
\begin{equation}
 m_{x} = \tanh \left[ (m_{x-1} + J_{x} m_{x} + m_{x+1} + h)/T\right]~,
\label{eq:mf}
\end{equation}
where $T$ is the temperature (we work in units where the Boltzmann constant is $k_B=1$). In the
clean case, $c=0$, all $J_x$ are equal to 1.  Consequently, all local mean-field equations are
identical, reading $m=\tanh[( 3 m +h )/T]$.   The clean critical temperature of the model
(\ref{eq:toy}) is therefore $T_c^0=3$.

In the presence of disorder, the local mean-field equations (\ref{eq:mf}) have to be solved
numerically. We employ a simple self-consistency cycle: All $m_x$ are initially set to $m_x=0.1$.
We then insert the current values of $m_x$ into the r.h.s.\ of (\ref{eq:mf}) and calculate new
values $m_x^{\rm new}$ which are then put into the r.h.s.\ and so on. This is repeated until the
change $|m_x^{\rm new} -m_x|$ falls below a threshold of $10^{-12}$ for all $x$. This method is
very robust, it never failed to converge, but it can be very slow in the tail region of the
transition where ordered islands coexist with large disordered regions. In this region sometimes
several 1000 iterations were necessary. However, in contrast to more sophisticated methods like
the Newton-Raphson method no matrix inversions are required.  Therefore we were able to simulate
very large systems up to $L_\bot=10^6$.

We also note that the model (\ref{eq:toy}) is similar to the mean-field McCoy-Wu model considered
in Ref.\ \cite{mfmccoy}. The raw data obtained in this paper ($L_\bot$ up to 1000) are very
similar to ours, but they were interpreted in terms of conventional power-law critical behavior
with an unusually large order parameter exponent $\beta\approx 3.6$. This may be partially due to
an unfortunate choice of parameters (in particular, a higher impurity concentration of $c=0.5$)
which makes it harder to observe the exponential tail of the smeared transition. We will come back
to this question in Sec.\ \ref{sec:discussion}.

\subsection{Thermodynamic magnetization and susceptibility}

\begin{figure}
\centerline{\includegraphics[width=\textwidth]{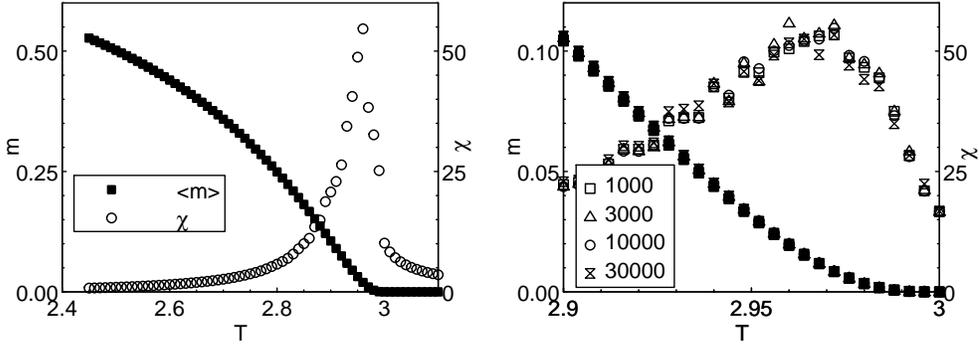}} \caption{Left: Magnetization
         $m$ and susceptibility $\chi$
         for $L_\bot=1000, ~c=0.2$. Right: Magnetization and susceptibility
         close to the seeming transition for different system sizes $L_\bot$ in
         the uncorrelated direction.} \label{Fig:magnetization_ab}
\end{figure}

In this subsection we present numerical results for the total magnetization $m=L_\bot^{-1} \sum_x
m_x$ and the homogeneous susceptibility $\chi = \partial m /\partial h$. The left panel of figure
\ref{Fig:magnetization_ab} gives an overview over the behavior of $m$ and $\chi$ as functions of
temperature $T$ for size $L_\bot =1000$ and impurity concentration $c=0.2$. The data are averages
over 1000 disorder realizations. (Thermodynamic quantities involve averaging over the whole
system. Thus ensemble averages rather than typical values give the correct infinite system
approximation.)
At a first glance these data suggest a sharp phase transition close to $T=2.96$. However, a closer
inspection of the transition region in the right panel of figure \ref{Fig:magnetization_ab} shows
that the singularities are rounded. If this smearing was a conventional finite-size effect the
magnetization curve should become sharper with increasing $L_\bot$ and the susceptibility peak
should diverge. This is not the case here, both the magnetization and the susceptibility data are
essentially size-independent. We conclude that the smearing is an intrinsic effect of the infinite
system.

\begin{figure}
\centerline{\includegraphics[width=\textwidth]{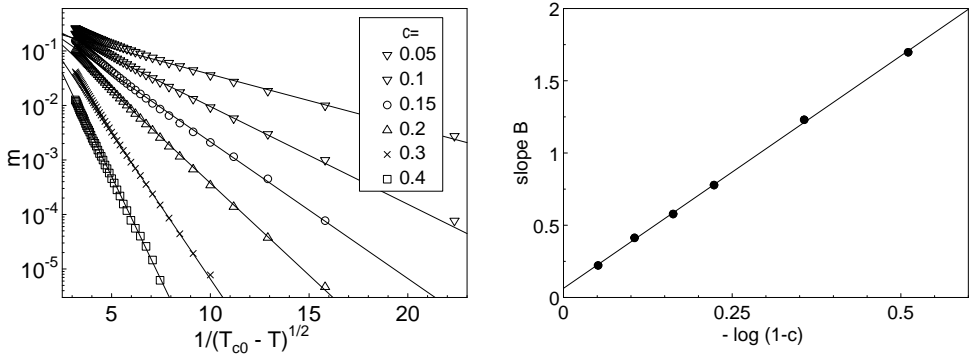}} \caption{Left:
         Total magnetization $m$ as a function of $(T_c^0-T)^{-1/2}$ for several impurity
         concentrations $c$.
         The solid lines is are fits to eq.\ (\ref{eq:m-dilute}) with $\phi=2$.
         Right: Decay constant $B$ as a function of $-\log(1-c)$.}
\label{Fig:log_comp}
\end{figure}

In order to compare the total magnetization data with the analytical Lifshitz tail prediction
(\ref{eq:m-dilute}) we need to determine the value of the clean finite-size scaling shift exponent
for the model (\ref{eq:toy}). Finite-size scaling is governed by the first appearance of a
non-vanishing solution of the linearized clean mean-field equation in a slab geometry of varying
thickness. This equation is equivalent to a one-dimensional Schr\"{o}dinger equation in a
potential well, leading to a clean shift exponent of $\phi=2$ and $d_\bot/\phi=1/2$.  In the left
panel of figure \ref{Fig:log_comp}, we therefore plot the logarithm of the total magnetization,
averaged over 300 samples, as a function of $(T_{c}^0-T)^{-1/2}$ for system size $L_\bot=10000$
and several impurity concentrations $c$.
For all $c$, the data follow eq.\ (\ref{eq:m-dilute}) over several orders of magnitude in $m$.
Fits of the data to (\ref{eq:m-dilute}) are used to determine the decay constants $B$. The right
panel of figure \ref{Fig:log_comp} shows that these decay constants depend linearly on
$-\log(1-c)$. This is exactly the expected behavior in a lattice model with binary disorder since
the probability for finding an island of size $L_R$ devoid of impurities behaves like $(1-c)^{L_R}
\sim e^{L_R\log(1-c)}$.

\subsection{Local magnetization}

\begin{figure}
\centerline{\includegraphics[width=\textwidth]{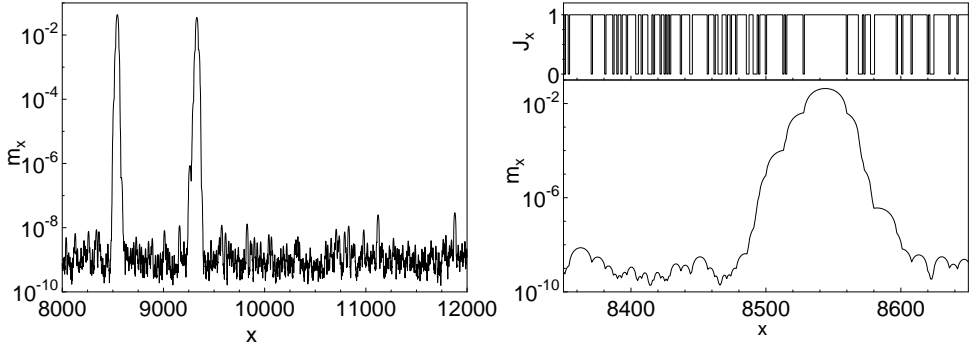}} \caption{Left:
         Local magnetization $m_x$ vs. $x$ for a segment of a single sample with size
         $L_\bot=100000$, impurity concentration $c=0.2$ and temperature $T=2.99$.
         Right: Local magnetization $m_x$ and local coupling constant $J_x$ in the vicinity of one
         of the ordered islands.}
\label{Fig:profile}
\end{figure}

In the tail region of the smeared transition the system consists of a few rare ordered regions (on
sufficiently large islands devoid of impurities). These rare regions are far apart, and between
them the local magnetization is exponentially small. This behavior is illustrated in figure
\ref{Fig:profile}.
The left panel shows the local magnetization of a segment of a large sample in the tail region of
the smeared transition. The data show that the local magnetization on the islands is large
($\approx 0.1$) but it drops very rapidly with increasing distance from the islands. This drop-off
can be used to estimate the bulk correlation length which is very small in this parameter region,
$\xi_0 \approx 3 ... 4$. Note that the seeming saturation of the magnetization decrease at
$m\approx 10^{-9}$ is a result of the finite external field $h=10^{-10}$. Without field, the
magnetization would drop much further. A comparison of the local magnetization $m_x$ and the local
coupling constant $J_x$ in the right panel of figure \ref{Fig:profile} shows that magnetic order
only exists on a sufficiently large island devoid of any impurities.

\begin{figure}
\centerline{\includegraphics[width=0.65\textwidth]{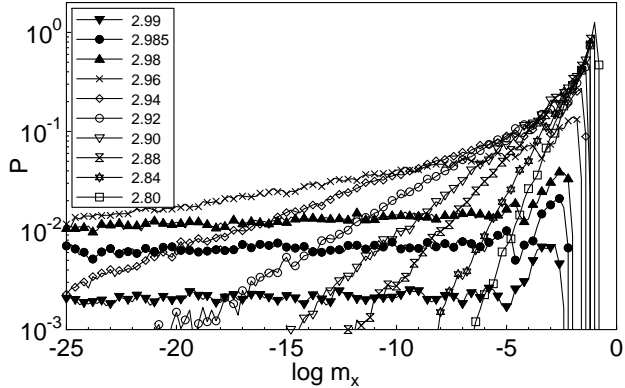}} \caption{Probability
distribution of the local magnetization $m_x$ for single sample of size $L_\bot=200000$, impurity
concentration $c=0.2$ and several temperatures. The three curves represented by filled symbols
($T\ge 2.98$) are in the asymptotic regime described by eq. (\ref{eq:plogm}).}
\label{Fig:mag_distrib}
\end{figure}

In order to quantify these observations, we have also calculated the probability distribution
$P(\log m_x)$ of the local magnetization values, or, more precise, their logarithms. Figure
\ref{Fig:mag_distrib} shows this distribution for a single large sample of $L_\bot=200000$,
impurity concentration $c=0.2$, and temperatures ranging from $T=2.8$ (deep in the ordered phase)
to $T=2.99$ (in the tail of the smeared transition).
Clearly, with increasing temperatures the distribution becomes very broad, even on this
logarithmic scale. Two qualitatively different regimes can be distinguished. For temperatures not
too close to the clean critical point ($T=2.80$ to 2.96), the distribution is dominated by an
exponential decay towards small magnetization values (i.e., towards negative $\log m$) with the
decay constant decreasing with increasing temperature. However, once the system enters the
rare-region dominated tail region of the smeared transition (which, as can be seen from figures
\ref{Fig:magnetization_ab} and \ref{Fig:log_comp}, happens roughly at $T=2.96 ... 2.97$ for
$c=0.2$) the behavior changes: In agreement with (\ref{eq:plogm}) the probability distribution
$P(\log m_x)$ becomes constant except for a peak at large magnetization values which we attribute
to the behavior of $m_x$ {\em on} the ordered islands (eq. (\ref{eq:plogm}) was derived from the
behavior of the sites {\em not} belonging to one of the rare ordered islands).

\subsection{Finite-size effects and sample-to-sample fluctuations}

In this subsection we present numerical data for the effects of a finite sample size $L_\bot$  in
the uncorrelated direction, i.e., perpendicular to the impurities. Effects of a finite size $L_C$
in the correlated directions cannot easily be studied within our approach because the limit $L_C
\to \infty$ has been taken in deriving the local mean-field equations (\ref{eq:mf}).

\begin{figure}
\centerline{\includegraphics[width=0.5\textwidth]{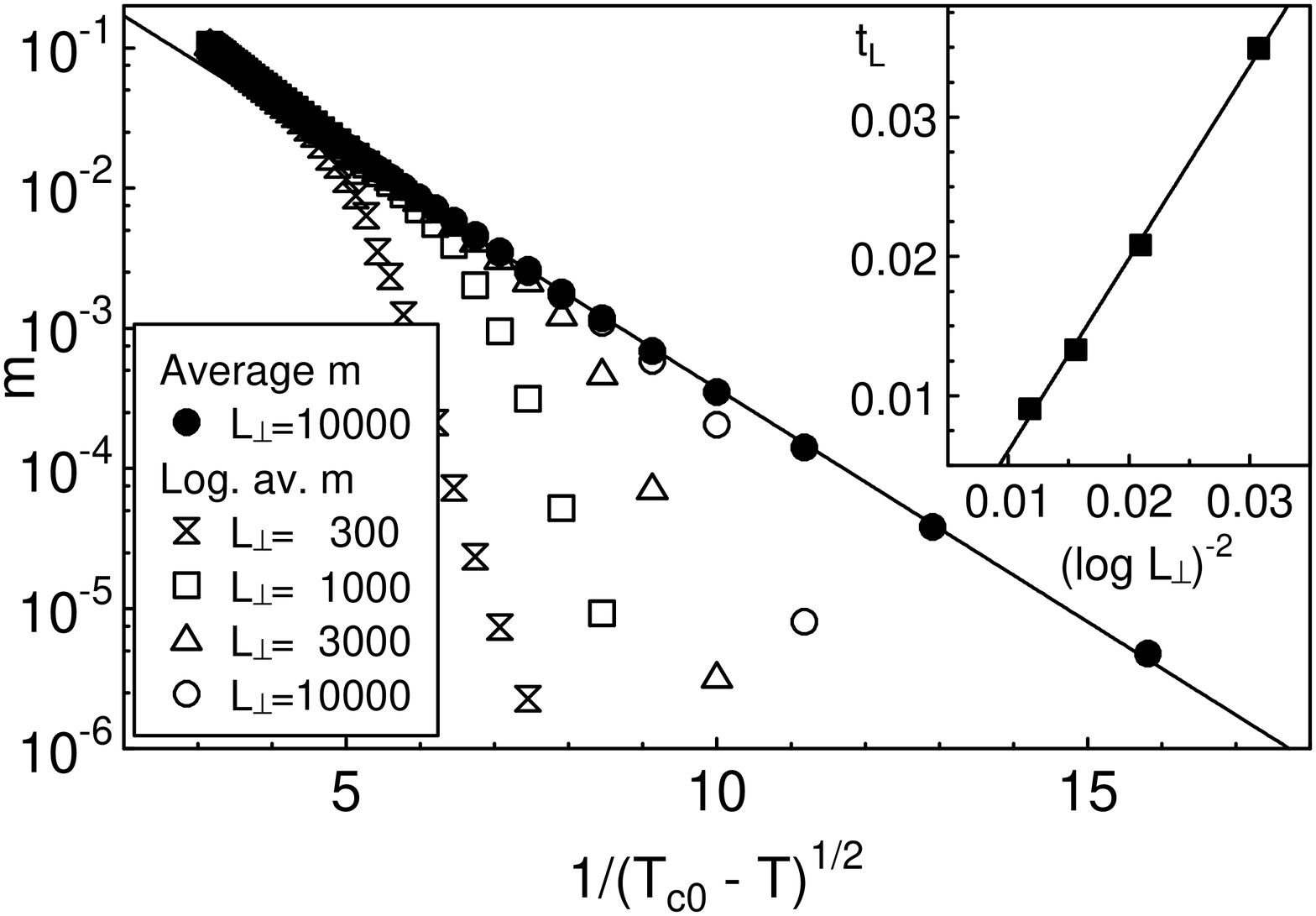}
             \includegraphics[width=0.5\textwidth]{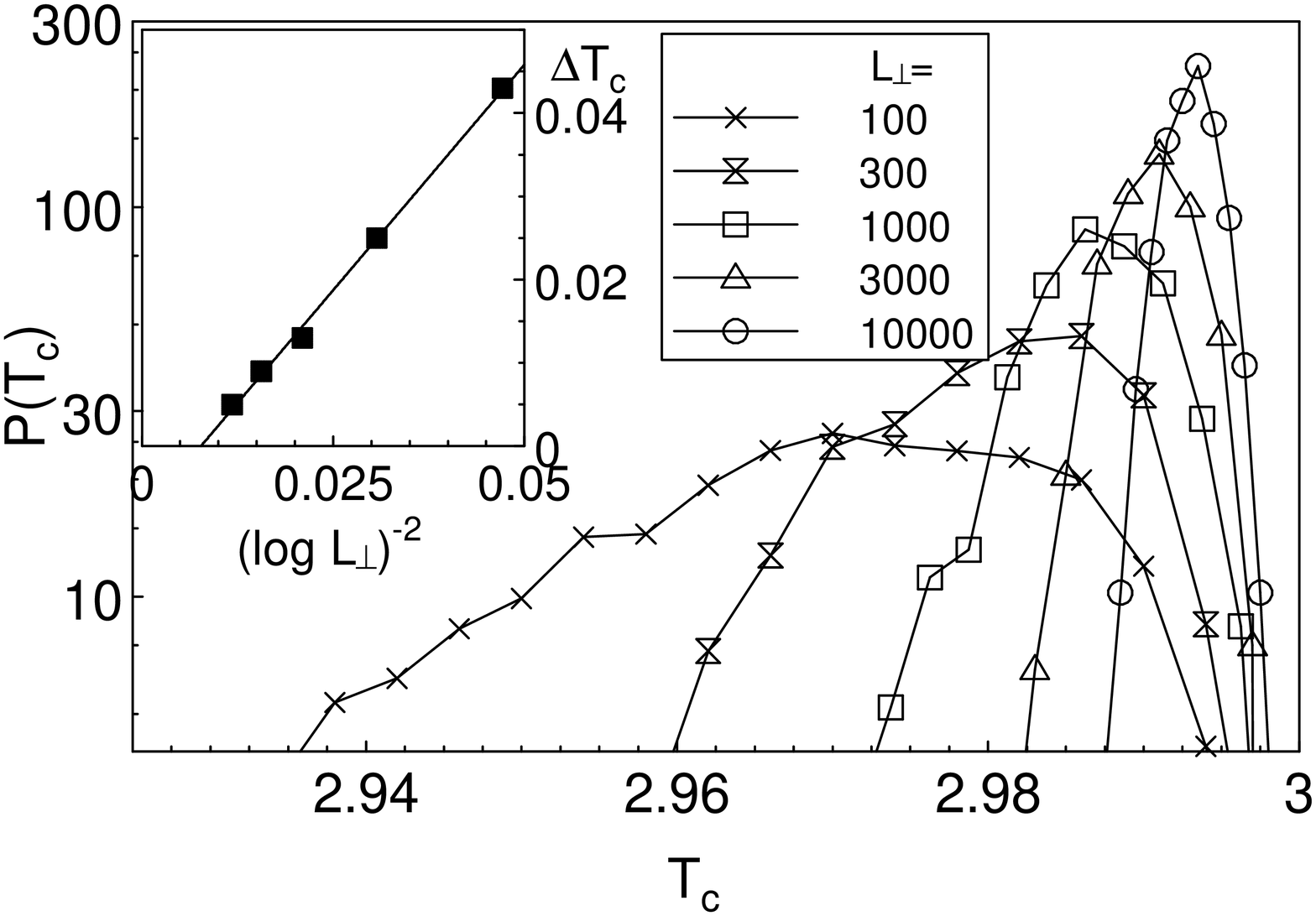}} \caption{Left: Average and typical
             (log. averaged) sample magnetization for $c=0.2$ and different sample sizes. The inset shows the
             onset temperature of strong sample-to-sample fluctuations (measured by
             $t_L=T_c^0-T_L$) as a function of $(\log L_\bot)^{-2}$. Right: Probability
             distribution of the sample ordering temperatures for $c=0.2$ and different sample
             sizes. The inset shows the width $\Delta T_c$ of this distribution as a function of $(\log
             L_\bot)^{-2}$.}
\label{Fig:finite_size}
\end{figure}

To investigate sample-to-sample fluctuations in a finite-size system we compare the average sample
magnetization $m_{av} = \langle m \rangle$ and the typical sample magnetization $m_{typ} = \exp
\langle \log m \rangle$ where $m$ is the total magnetization of a sample and  $\langle \cdot
\rangle$ denotes the average over the disorder. A significant difference between average and
typical magnetizations indicates strong sample-to-sample fluctuations. The left panel of figure
\ref{Fig:finite_size} shows the typical magnetizations for four different sizes $L_\bot$ together
with the average magnetization (which is size-independent within the accuracy of this plot) for
300 to 1000 samples, depending on $L_\bot$.
In the tail region of the smeared transition the typical and average magnetizations deviate from
each other by several orders of magnitude. From these data we define a temperature $T_{L_\bot}$
which describes the onset of the sample-to-sample fluctuations via $m_{av}(T_{L_\bot}) /
m_{typ}(T_{L_\bot}) = \rm e$. This temperature shifts to larger values with increasing system size
$L_\bot$. The inset shows $t_{L_\bot} = T_c^0 -T_{L_\bot}$ as a function of $(\log L_\bot)^{-2}$.
The plot shows that $t_{L_\bot}$ follows  eq.\ (\ref{eq:tL-dilute}) in good approximation.

The right panel of figure \ref{Fig:finite_size} shows the probability distribution of the sample
critical temperature which is defined as the ordering temperature of the largest rare region in a
particular sample (600 to 2000 samples used).  As predicted in subsection \ref{subsec:FS}, the
distribution becomes narrower with increasing size $L_\bot$, and it shifts towards the clean
critical temperature $T_c^0=3$. The width $\Delta T_c$ of this distribution has been determined
from the values where $P$ has decreased by a factor of 1/e. The inset shows that $\Delta T_c$ is
proportional to $(\log L_\bot)^{-2}$ as predicted in eq.\ (\ref{eq:dtc}).

\section{Discussion}
\label{sec:discussion}

To summarize, we have shown that true static order can develop on an isolated rare region in an
Ising model with planar defects. As a result, different parts of the system undergo the phase
transition at different temperatures, i.e., the sharp transition is smeared by the defects. In
this final section we discuss the generality of our findings and their relation to Griffiths
phenomena and the Harris criterion. We also discuss favorable conditions for observing the
smearing in experiments or simulations.

The origins of the smearing of the phase transition introduced in this paper and of Griffiths
phenomena are very similar, both are caused by rare large spatial regions which are locally in the
ordered phase. The difference between the two effects is a result of disorder correlations. For
uncorrelated or short-range correlated disorder, the rare regions are of finite size. Thus, they
cannot develop true static long-range order. Instead, the order parameter fluctuates slowly
leading to the Griffiths singularities \cite{Griffiths} discussed in Sec.\ \ref{sec:intro}. In
contrast, if the rare regions are infinite in at least two dimensions, a stronger effect occurs:
the rare regions which are locally in the ordered phase actually develop true static order, and
this leads to a smeared transition. In other words, the same rare regions which would usually
produce Griffiths singularities are responsible for the smearing of the transition in a system
with planar defects. The tail of the smeared transition stretches to where the Griffiths
temperature would normally be, i.e., to $T=\infty$ for Gaussian disorder or $T=T_c^0$ for Poisson
disorder. So, the smeared transition replaces not only the conventional critical point but also
the Griffiths phase.

We now discuss under which conditions a phase transition will be smeared by disorder. The
arguments given in Sec.\ \ref{sec:rounding} in favor of a smeared transition only relied on the
dimensionality of the rare regions and the order parameter symmetry. It is therefore expected that
all phase transitions with discrete order parameter symmetry are smeared by quenched disorder if
the defects are perfectly correlated in at least two dimensions. In systems with linear defects
the transition will remain sharp if the interactions are short-ranged. However, if the
interactions in the correlated direction fall of as $1/r^2$ or slower, even linear defects can
lead to smearing because a 1D Ising model with $1/r^2$ interaction has an ordered phase
\cite{Ising1r2}. This is particularly important \cite{rounding_prl} for quantum phase transitions
in itinerant electronic systems which can be mapped onto classical systems with $1/\tau^2$
interaction in imaginary time direction.

Phase transitions with continuous order parameter symmetry are more stable against smearing. In
systems with short-range interactions the dimensionality of the defects has to be at least three,
and in systems with linear or planar defects long-range interactions are required for smearing. It
is known \cite{bruno01} that classical XY and Heisenberg systems in dimensions $d=1,2$ develop
long-range order at finite $T$ only if the interaction falls off more slowly than $1/r^{2d}$.
Consequently, in a system with linear defects the phase transition will only be smeared if the
interactions in defect direction fall off more slowly than $1/r^2$ (or more slowly than $1/r^4$
for planar defects).

A third important question concerns the universality of the thermodynamic behavior in the vicinity
of the smeared transition. As can be seen from comparing the results of Poisson and Gaussian
disorder, the functional dependence of the magnetization and other observables on the temperature
is {\it not} universal, it depends on details of the disorder distribution. Therefore, only the
presence or absence of smearing is universal in the sense of critical phenomena (i.e., depending
on dimensionality and symmetry only) while the thermodynamic relations are non-universal.

Our fourth remark deals with the relation of the disorder-induced smearing and the Harris
criterion which in the case of extended defects reads \cite{BoyCardy} $\nu > 2/d_\bot$. We
emphasize that the phase transition can be smeared by planar defects even if the corresponding
clean critical point fulfills the Harris criterion and appears to be stable. The reason is that
the Harris criterion assumes a {\em homogeneous} transition and studies the behavior of the
coarse-grained (root-mean-square) disorder at large length scales. However, the formation of
static order on an isolated finite-size rare region is a non-perturbative finite-length scale
effect in the tail of the disorder distribution. This type of effects is not covered by the Harris
criterion.

We now turn to the question which conditions are favorable for the {\it observation} of the
smearing in experiments or simulations. A significant number of large rare regions will only exist
if the defect concentration is small, as can be seen  from (\ref{eq:w-dilute}). On the other hand,
the impurities have to be sufficiently strong so that the bulk system is still far away from
criticality when the first rare regions start to order. Thus, the most favorable type of disorder
for the observation of the smearing is a small density of strong impurities. In contrast, for weak
impurities the rounding is restricted to a very narrow temperature range below the clean critical
temperature and maybe masked by the remainder of the bulk critical fluctuations. If the impurity
concentration is too high, the exponential drop-off is very fast making it very hard to observe
the smearing. This may also explain why no smearing was observed in earlier simulations.
Specifically, in Ref.\ \cite{LeeGibbs} weak impurities ($\Delta J/J=0.1$) of a high concentration
$c=0.5$ were used. These are unfavorable conditions, and the maximum system size of $L=27$ used in
this simulation was probably too small to observe the smearing.

Finally, we add one remark about the extreme tail of the smeared transition. In this parameter
region, the ordered islands are very far apart and the aligning effect of the direct island-island
interactions is very weak. It is therefore possible that in a real system other interactions play
an important role in determining the relative orientation of the order parameter on the islands.
This could lead, e.g., to spin glass order in the extreme tail of the smeared transition.

In conclusion, perfect disorder correlations in one or more directions dramatically increase the
effects of the disorder on a phase transition with discrete order parameter symmetry.
Zero-dimensional defects typically lead to a conventional critical point with power-law scaling.
For linear defects the generic behavior seems to be an infinite-randomness critical point
\cite{McCoy69,dsf9295,Motrunich00} while we have shown here that planar defects generically
destroy the sharp transition by smearing.

\ack We acknowledge support from the University of Missouri Research Board. Part of this work has
been performed at the Aspen Center for Physics.
%
%


\section*{References}
\begin{thebibliography}{99}
\frenchspacing
\bibitem{Grinstein} Grinstein G 1985 {\it Fundamental Problems in Statistical
 Mechanics VI}, ed E G D Cohen (New York:Elsevier) p.147
\bibitem{Harris74} Harris A B 1974 {\it J. Phys. C} {\bf 7} 1671
\bibitem{AharonyHarris96} Aharony A and Harris A B 1996 {\it Phys. Rev. Lett.} {\bf 77} 3700
\bibitem{WisemanDomany98} Wiseman S and Domany E 1998 {\it Phys. Rev. Lett.} {\bf 81} 22
\bibitem{McCoyWu} McCoy B M and Wu T T 1968 {\it Phys. Rev.} {\bf 176} 631
\par\item[] McCoy B M and Wu T T 1969 {\it Phys. Rev.} {\bf 188} 982
\bibitem{Griffiths} Griffiths R B 1969 {\it Phys. Rev. Lett.} {\bf 23} 17
\bibitem{Randeria} Randeria M, Sethna J, and Palmer R G 1985 {\it Phys. Rev. Lett.} {\bf 54} 1321
\bibitem{Bray89} Bray A J and Huifang D 1989  {\it Phys. Rev. B} {\bf 40} 6980
\bibitem{Dhar} Dhar D 1983 {\it Stochastic Processes: Formalism and Applications}, ed D S Argawal
   and S Dattagupta (Berlin:Springer)
\bibitem{Dhar88} Dhar D, Randeria M, and Sethna J P 1988 {\it Europhys. Lett.} {\bf 5} 485
\bibitem{Bray88a} Bray A J 1988 {\it Phys. Rev. Lett.} {\bf 60} 720
\bibitem{Bray88b}Bray A J and Rodgers G J 1988 {\it Phys. Rev. B} {\bf 38} 9252
\bibitem{Bray87} Bray A J 1987 {\it Phys. Rev. Lett.} {\bf 59} 586
\bibitem{Dreyfus} Dreyfus H von, Klein A, and Perez J F 1995 {\it Commun. Math. Phys.} {\bf 170} 21
\bibitem{Gielis95} Gielis G and Maes C 1995 {\it J. Stat. Phys.} {\bf 81} 829
\bibitem{Cesi97} Cesi S, Maes C, and Martinelli F 1997 {\it Commun. Math. Phys.} {\bf 189} 135
\par\item[]Cesi S, Maes C, and Martinelli F 1997 {\it Commun. Math. Phys.} {\bf 189} 323
\bibitem{Lubensky} Lubensky T C 1975 {\it Phys. Rev. B} {\bf 11} 3573
\bibitem{Rudnick} Rudnick J 1978 {\it Phys. Rev. B} {\bf 18} 1406
\bibitem{Andelman} Andelman D and Aharony A 1985 {\it Phys. Rev. B} {\bf 31}
    4305
\bibitem{Dorogovtsev} Dorogovtsev S N 1980 {\it Fiz. Tverd. Tela (Leningard)} {\bf 22}
    321 [{\it Sov. Phys.--Solid State} {\bf 22} 188]
\bibitem{BoyCardy} Boyanovsky D and Cardy J L 1982 {\it Phys. Rev. B} {\bf 26} 154
\bibitem{DeCesare} De Cesare L 1994 {\it Phys. Rev. B} {\bf 49} 11742
\bibitem{LeeGibbs} Lee J C and Gibbs R L 1992 {\it Phys. Rev. B} {\bf 45} 2217
\bibitem{McCoy69} McCoy B M 1969 {\it Phys. Rev. Lett.} {\bf 23} 383
\bibitem{dsf9295} Fisher D S 1992  {\it Phys. Rev. Lett} {\bf 69} 534
\par\item[] Fisher D S 1995 {\it Phys. Rev. B} {\bf 51} 6411
\bibitem{YoungRieger96}Young A P and Rieger H 1996 {\it Phys. Rev. B} {\bf 53} 8486
\bibitem{Pich98}Pich C, Young A P, Rieger H, and Kawashima N 1998 {\it Phys. Rev. Lett.} {\bf 81} 5916
\bibitem{Motrunich00} Motrunich O, Mau S-C , Huse D A, and Fisher D S 2000 {\it Phys. Rev. B} {\bf 61} 1160
\bibitem{ma79} Ma S K, Dasgupta C, and Hu C-K  1979 {\it Phys. Rev. Lett.} {\bf 43} 1434
\bibitem{bhattlee82} Bhatt R N and Lee P A 1982 {\it Phys. Rev. Lett.} {\bf 48} 344
\bibitem{fisher94} Fisher D S 1994 {\it Phys. Rev. B} {\bf 50} 3799
\bibitem{huse_private}Huse D A private communication
\bibitem{Lifshitz}Lifshitz I M 1964 {\it Usp. Fiz. Nauk} {\bf 83} 617
                  [{\it Sov. Phys.--Usp.} {\bf 7} 549]
\par\item[] Friedberg R and Luttinger J M 1975 {\it Phys. Rev. B} {\bf 12} 4460
\bibitem{HalperinLax}Halperin B I and Lax M 1966 {\it Phys. Rev.} {\bf 148} 722
\bibitem{Sknepnek} Sknepnek R and Vojta T unpublished
\bibitem{karevski}Karevski D and Turban L 1996 {\it J. Phys. A} {\bf 29} 3461
\bibitem{mfmccoy}Berche B, Berche P E, Igloi F, and Palagyi G 1998 {\it J. Phys. A} {\bf 31} 5193
\bibitem{Ising1r2}Thouless D J 1969 {\it Phys. Rev.} {\bf 187} 732
\par\item[]Cardy J 1981 {\it J. Phys. A} {\bf 14} 1407
\bibitem{rounding_prl}Vojta T 2003 {\it Phys. Rev. Lett.} {\bf 90} 107202
\bibitem{bruno01}Bruno P 2001 {\it Phys. Rev. Lett.} {\bf 87} 137203

\end{thebibliography}
\end{document}